December 11, 2018

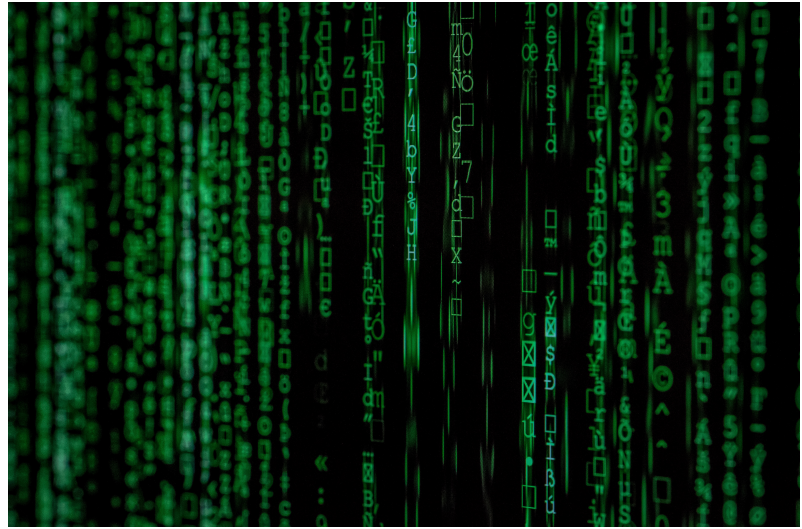

# Security and Privacy Implications of Middlebox Cooperation Protocols

**White paper by the EU-H2020 MAMI project (grant agreement No 688421)**

Thomas Fossati (Nokia), Roman Müntener (Zurich University of Applied Sciences),
Stephan Neuhaus (Zurich University of Applied Sciences), and Brian Trammell (ETH Zurich)

The Measurement and Architecture for a Middleboxed Internet (MAMI) project is an EU-H2020 funded project that started in January 2016. Further information about the project can be found on the [MAMI project website](#).

**This white paper presents an analysis done by the MAMI project of the privacy and security concerns surrounding middlebox cooperation protocols (MCPs), based on our experimental experience with the Path Layer UDP Substrate (PLUS) proposal. Our key finding is that adding explicit signaling meant for on-path devices presents no significant new attack surface as compared to the status quo in the Internet architecture. While middlebox cooperation can make a passive adversary's job easier, it does not enable entirely new attacks.**

One of the main goals of the MAMI project is to (re-)enable innovation in the transport layer and to end ossification of transport protocols. Based on the end-to-end principle, the Internet was originally envisioned to consist of "smart endpoints and dumb pipes," where the intermediates only know how to move an IP packet toward its destination, nothing more. However, not all routers present in the Internet are "dumb", having been augmented by middleboxes that do all kinds of things with transports.

This is problematic because these middleboxes must now look at the transport headers or higher layers in order to perform their function. And that means that they have to know what the transport protocol is, because otherwise they won't know where to look. This means that the choice of transport is very restricted, and in most cases, that transport will either be TCP or alternatively UDP with a specific transport layered on top of it. If a packet using a different transport comes along, often a middlebox wouldn't know what to do with it and may well drop it. This leads to the entrenchment of TCP, or ossification of transport protocols.

Another development is ubiquitous encryption on the Internet. This makes the job of some middleboxes harder because they may not be able to look far enough into a packet in order to make a decision. The solution in many cases is to violate the end-to-end security property that is nominally guaranteed by protocols such as TLS so that the middleboxes can do their work. This is an unsatisfactory solution for some. However, reinstating end-to-end encryption including transport headers would make many middleboxes blind.

*With a path layer in place, the transport layer can be encrypted while a middlebox can do its work, without being aware of the specifics of the transport protocol in use. But does this approach pose privacy and security issues worse than the problems it solves?*

The solution proposed and developed within the MAMI project is to make certain information explicit in a new [path layer](#) [1] as a shim between the network and transport layers, designed explicitly for per-flow communication with stateful on-path devices. The information exchanged in the path layer would, for example, include "is this the first packet in a flow?", "is this the last packet in a flow?", "is this packet more loss-sensitive or more latency-sensitive?", "is there congestion on the path that this packet traveled on", and so on. With the path layer in place, the transport layer, including headers, could be encrypted while a middlebox could do its work without being aware of the specifics of the transport protocol in use. This would solve both problems: transport innovation is possible again because the precise location and meaning of transport-specific information is no longer needed; and since the information that a middlebox needs is now provided outside the encrypted transport, they are no longer blind.

In this white paper we address the question of whether such middlebox cooperation protocols (MCP) potentially pose security and privacy issues outweighing the problems that they solve, as:

a.  MCPs expose information explicitly that adversaries would normally have to work for, and

b.  MCPs might enable completely new attacks on user privacy.

Of these, (a) is in fact true. Indeed the entire point of an MCP is to explicitly expose information that on-path elements would otherwise have to work for. While the MCP is designed for beneficial middleboxes, what holds for them must also be true for adversaries. MCPs are therefore at best neutral to privacy. However, our analysis also indicates that (b) is not true. This means that the question of whether or not to employ an MCP now boils down to risk analysis: do the benefits of MCPs balance the increased ease with which passive adversaries can obtain information?

To examine this question, we first consider a generalized model for any protocol that exposes information for middlebox cooperation, and then a threat model for misusing both the information exposed and the mechanisms used to expose it.

# A Model of Middlebox Cooperation

A middlebox cooperation protocol provides mechanism to communication information to the middlebox on the path or mechanism for middlebox on the transmission path to communicate information to the endpoints. Our model middlebox cooperation protocol is based on the Path Layer UDP Substrate (PLUS) protocol described in detail in our 2017 CNSM paper "[A Path Layer for the Internet: Enabling Network Operations on Encrypted Protocols](#)" [1]. PLUS is a dedicated protocol to signal from and to middleboxes between an encrypted transport layer and the network layer protocol.

Based on the analysis of PLUS, our model exposes four types of information to on-path devices:

- Identification: for the assignment of packet to a flow. All stateful in-network devices need to identify a packet belong to a flow as a first step in order to perform the desired network function.

- Flow state: the establishment of a connection between two endpoints, as well as the maintenance and disestablishment of that connection. A provision for in-network flow state is necessary today due to the prevalence of stateful network address translators (NATs), bearers and tunnels, and stateful firewalls in the Internet and often derived from the cleartext information in the TCP header; lack of flow state signaling leads to the necessity of unproductive keepalive traffic to maintain flow state based on idle timeouts.

- Signaling for packet treatment desired by the endpoint, in terms of application demand and transport layer characteristics.

- Measurement and measurability, supporting on-path estimation of basic traffic metrics equivalent to unencrypted TCP, such as re-ordering, latency, or loss.

In our model MCP this information is signaled using following key types of metadata that are assumed to be directly exposed in the (path layer) protocol *headers*:

- A *connection identifier* identifying the flow to which a packet belongs

- A *packet number* identifying a packet within the sequence of packets in a flow, and to support on-path upstream loss measurement

- A *packet number echo* used to demonstrate receipt of a packet in the opposite direction, and to support two-way latency measurement

- State signals for signaling flow start and/or flow end, to support on-path state maintenance.

- *Treatment signals* for signaling the tradeoffs a sender is willing to make for network treatment of a packet.

In PLUS there is a one-to-one mapping between these types of metadata and header fields or flags. Each of these header fields has predefined semantics in the protocol specification and cannot be modified undetected by on path devices.

Note that much of this signaling is today provided implicit and explicit transport- and network-layer signals; for example, the

*Much of the signaling provided by middlebox cooperation protocols that do not expose payload is provided by implicit and signals in today's network stack. Our analysis is therefore informed by looking at information radiated in the headers of current, non-encrypted protocols.*

TCP wire image provides for flow state maintenance (through the TCP flags) and measurability (via sequence and acknowledgment numbers as well as timestamps).

Our model MCP also provides a facility for an on-path device to explicitly send small signals to both the sender and a receiver of a flow, for example to identify itself as a firewall and to advertise its policy, or to assist in the measurement of the maximum transmission unit (MTU) along the path. This facility requires endpoint permission for the on-path device to send a signal, and uses (encrypted) feedback from the receiver back to the sender to close the loop (i.e., to allow the middlebox to signal not just to the packet's receiver, but also its sender).

Signaling by on-path middleboxes uses typed, constant-length *scratch space* allocated in the packet by the sender. These on-path middleboxes are prevented from making undetected changes to the rest of the packet by cryptographic integrity protection of the entire header except the content of the scratch space, while protecting the length and type of the scratch space. In contrast to header fields, scratch space has a representation and semantics varying according to the values of other header fields, and may be writable by middleboxes if allowed (via the integrity protection mechanism) by the endpoints.

Key to understanding this model is that MCPs provide a cryptographically-reinforced boundary among three classes of information:

- information that is solely in the *end-to-end trust domain* (i.e., for the use of the endpoints and any devices with which those endpoints share their cryptographic keys);

- information that can be inspected by devices outside the end-to-end trust domain, but not modified;

- information that can be inspected and modified by devices outside the end-to-end trust domain.

While this model is derived directly from our experimentation with PLUS, other approaches to limited path cooperation (such as the TLS or QUIC connection IDs, or the QUIC spin bit) expose some or all of the metadata PLUS does. To our knowledge, all approaches in this space fit this model, though there are some which operate primarily through selective expansion of the end-to-end trust domain. We therefore consider our analysis to be generalizable.

# Attacker Model

We build upon the attacker model for a pervasive passive observer outlined in RFC 7624 [2], "Confidentiality in the Face of Pervasive Surveillance". We draw additional inspiration from Detecting and Defeating TCP/IP Hypercookie Attacks, [3], which examined side-channels in existing transport protocols.

We call the endpoint that initiates communication the *client*, and the other endpoint the *server*. We call the endpoint that sends a packet the *sender*, and the other endpoint the *recipient*.

Generalizing from the MCP model above, we assume the potential presence of header fields and scratch space, with varying levels of integrity protection. The headers and scratch space of a MCP must be exempt from encryption because they exist to communicate to middleboxes with whom the endpoints have no cryptographic association.

We assume that communication is encrypted and integrity protected by default at or above the transport layer, because this is increasingly the case. We will explicitly identify any attack that requires plaintext communication. By default the only information that an attacker can extract is metadata, and it has long been known that metadata can be used to infer information about the data.

Out-of-band identification methods, e.g., linking a flow's five- or six-tuple with an identifier and using some other protocol to export this linkage, are also not considered, because it is practically impossible for users and remote endpoints to detect and defeat. We say metadata is *exposed* when it can be either recorded or inferred.

Taking RFC 7624 [2] as the basis of our attacker model, we consider passive adversaries that are however allowed *some*

modification of the packet, if this furthers its goals. In summary, our attackers:

- "can observe every packet of all communications at any hop in any network path between the endpoints (this means the client cannot use anonymisation services like Tor)";

- "can observe data at rest in any intermediate system between the endpoints controlled by the endpoints";

- "can share information with other such attackers"; and

- may take other actions with respect to these communications (e.g., blocking, modification, injection, etc.), as long as these actions do not cause the communication to be totally disrupted.

In contrast to RFC 7624, we also consider an attacker that can also perform *active attacks* (short of denial of service). Blocking actions that fall short of denial of service include:

- dropping a single packet to force retransmission by the sender's transport layer;

- blocking requests to certain DNS servers to force use of other DNS servers;

- blocking TLS-protected communications to force unprotected communications;

- shutting down a selected number of links in a multi-path scenario and thus forcing the multi-path transport to use specific links, on which surveillance may be easier.

We stipulate that our attackers want to:

- remain undetected, if possible; and

- extract as much information as possible about the communication between the endpoints.

The attacker generally wants to remain undetected if at all possible, but this requirement might be relaxed if the consequences of detection are outweighed by the usefulness of the attack. This can happen for example if the attacker is so powerful that detection is without consequence (low risk), or if the attack, if successful, would give the attacker information of extremely high value (high pay-off).

## Classes of Manipulation

At a very high level, we may distinguish between two different and orthogonal features of manipulation:

- Manipulation that is detectable by the endpoint. For example, changing information that has been integrity-protected. We call this class of manipulation "class D".

- Manipulation that changes the observed protocol behaviour. For example, by dropping single packets to force packet retransmission. We call this class of manipulation "class P".

These classes are orthogonal because there are manipulations that fall into any of the four combination of class. For example, changing information that is integrity protected falls into class D, but may or may not fall into class P. Dropping packets falls into classes !D and P (if the protocol does retransmission), while changing the data in a packet and then changing it back before delivering it to the remote point falls into !D and very probably also into !P.

Since a mostly passive attacker wants to remain undetected, attacks that are in class D are undesirable, and attacks in class P may or may not also be undesirable, depending on whether the observed change in protocol behaviour is advantageous for the attacker or not.

Purely passive attacks are by necessity in class D, and attackers will prefer class (!D, !P) over (!D, P) over (D, *).

# On-Path Attacks Against an MCP

We first consider the actions available on an MCP to an on-path attacker following our attacker model above and then also examine the privacy and security implications of each type of information included in an MCP, on a per type basis (identification, state, and treatment) as whole as a whole (for fingerprinting). MCPs need only concern themselves with these attacks when they support the specific mechanisms or information considered.

## Data Exfiltration via Header Fields

Our model MCP cooperates with its overlying transport protocol to provide integrity protection for certain parts of its header or payload integrity. However, to be effective, integrity protection has to be checked and acted upon. If the MCP's specification does not mandate that integrity protection failures should cause hard failure of a connection, an attacker could manipulate header fields to cause different treatment by downstream middleboxes. On the other hand, it would be inappropriate for an explicit path layer protocol, like PLUS, to decide unilaterally that packets that fail the integrity check MUST be dropped. It may be that the packet payload carries identifying and authenticating information outside of a crypto context, and it may be that in certain circumstances, availability is more valuable than confidentiality. In these cases, it is perfectly acceptable to leave the decision whether to accept or reject a packet to the overlaying transport.

It is not completely clear how these header fields could be usefully exploited by an attacker acting at a single point in the network, since metadata insertion or data exfiltration is complicated by at least two factors:

- The amount of metadata per packet is limited (often on the order of a few bits) and exposed metadata should in general be minimal with respect to the use case the information is exposed for;

- Changing the bits might result in unwanted outcomes (class P).

These complicating factors are mitigated if the attacker can control both sides of the connection; in this case, the attacker can rewrite any header field which does not interfere with end-to-end transmission and, as long as the changes are undone before the packet reaches the remote endpoint, the attacker can exfiltrate data to the network portion between the attacker's ingress and egress nodes. This can be done in a way that is undetectable by either endpoint. In a more generic case, this type of attack requires a side channel for coordination between the attacker's nodes.

*An attacker that can control both ends of a segment of a network between two endpoints can exfiltrate information freely on communications between those endpoints, whether an MCP is in use or not.*

However, this mode of attack is not specific to middlebox cooperation protocols: the present protocol stack has many header fields that can be practically exploited in this way even without middlebox cooperation (e.g., IPv4's IPID on non-fragmenting paths, IPv6's Flow ID). As these are often unauthenticated, two-point control might not even be necessary. For an exhaustive investigation of this kind of attack, see also [Detecting and Defeating TCP/IP Hypercookie Attacks](#) [3].

For integrity-protected headers, this attack is in class (D, *), and hence not preferred. For other headers, this attack is in class (!D, *), and very probably in class (!D, !P), and hence much preferred.

## *Data Exfiltration Through Scratch Space*

For payload data that is not integrity-protected, such as scratch space, exfiltration becomes a much more attractive option, since manipulation of scratch space is potentially undetectable by the endpoint. But even here, an attacker must exercise some care not to trigger unwanted (to the attacker) behaviour by manipulating the scratch space. The utility of this attack is starkly limited when (as in our model MCP) the size and type of the scratch space is integrity protected by the receiving endpoint, and the size and type (indeed, the very presence) of the scratch space is under the sending endpoint's control.

This attack is in class (!D, *), and therefore desirable. However, it is an active attack that might not be preferred by a mostly passive adversary.

## *Coercion of Scratch Space*

Further, access networks could require endpoint owners to supply packets with specified scratch space or otherwise to refuse to forward the packets, or to refuse to forward them speedily. This would follow the familiar pattern in which opt-in solutions to data tracking decay to a more mandatory position, as companies only offer certain services to users who do in fact opt in. People who choose not to opt in are unable to use this service. If there is one dominant service of its kind available (e.g., Facebook or Twitter), or if there is only one viable alternative, people may find they have no choice and need to opt in.

Access network providers or even core networks could do the same to ease data exfiltration. They could force endpoints to provide scratch space, or face the penalty of being dropped or of being put into the slow lane if they do not comply.

This attack is in class (D, !P), and also active, and would therefore not be preferred by the mostly passive adversary.

## Connection Identification and Linkability

Identification and linkability of packets to connections or across connections is possible through any number of constructs in the packet header. Here we consider two types of linkable identifier, as provided in our model MCP: connection identifiers and (echoed) packet numbers.

A connection identifier (CID) is a transport layer construct that allows endpoints and middleboxes to uniquely identify an end-to-end connection/session even if the underlying 5-tuple changes due to NAT re-bindings, connection migration with multi-homing or multi-path. Examples of middleboxes that could utilise this information include load balancers and firewalls.

The existence of an explicit CID is especially important in association with encrypted protocols where session re-negotiation involves a handshake that is expensive in terms of computation and latency. The presence of a CID does not affect the overall security of the session with respect to authentication, confidentiality and integrity. The intrinsic properties of the CID, though, have obvious privacy implications. Namely, the ability of a passive on-path attacker to relate packets belonging to the same logical flow (and therefore determine whether two or more items of interest are coupled), this introduces *linkability* (see [RFC6973](RFC6973)). Therefore, a good CID design should take this into account and seek, as much as possible, to avoid linkability by passive on-path adversaries between multiple source addresses / ports during mobility or NAT rebind scenarios. This may not be always possible - e.g., when clients are unaware of the address change (for example, when passing through NATs.) An example of a design that satisfies this requirement is the HOTP-based construction in [draft-mavrogiannopoulos-tls-cid](draft-mavrogiannopoulos-tls-cid).

Packet numbers and echoes (as provided in our model MCP, and ubiquitously present in Internet traffic in a different form, as TCP sequence and acknowledgment numbers) are also usable for linkability. From the passive observer's point of view, they are a "poor man's connection identifier". In this discussion, we assume that packet numbers start with a random initial number in a given number space and are incremented for each new packet sent, as in TCP and the PLUS PSN. An attacker may assume that a packet seen with a packet number that is not within a small delta of other known packet numbers probably belongs to a new connection. Likewise, a packet seen by the attacker that *is* within a small delta, but has a different source IP, probably indicates a change of the endpoint source IP address. These conclusions, however, come with some uncertainty, so PSN information is in this sense less reliable than CID from the attacker's point of view.

An important security property of a CID or initial PSN is that it should be unpredictable and difficult for an off-path attacker to guess, for example to prevent an adversary to blindly reset a transport session or confuse a firewall about the state of its monitored flows.

There are multiple possible design options for implementing connection identifier selection (server picks, client picks, negotiated, mono or bi-directional - see e.g. PLUS, QUIC, draft-rescorla-tls-dtls-connection-id, draft-mavrogiannopoulos-tls-cid, draft-barrett-mobile-dtls, IPsec). The one that is emerging with QUIC gives the server the ability to pick the CID that best suits the server. This is to allow load balancers to work cooperatively with servers by using the CID to identify which server to send traffic towards, preferably in a stateless manner. There are caveats to be aware of when using this approach:

- The opportunity for selective DoS, i.e., overloading a particular backend server behind the load balancer. This could be avoided / mitigated by including a lightweight authenticator that the load balancer can use to drop CID values that don't validate;

- A privacy problem related to the potential leakage of topological information about the server farm.

Attacks against CID and PSN/PSE are in class (!D, !P), but there are privacy-preserving designs for connection identification, as well.

## Information Exposure of State Signals

The state signals in our MCP model are designed to signal start of flow and end of flow. These signals can drive an abstract, transport-independent state machine shown in the figure here; however, this state machine is merely one instance of a set of on-path state maintenance behaviors.

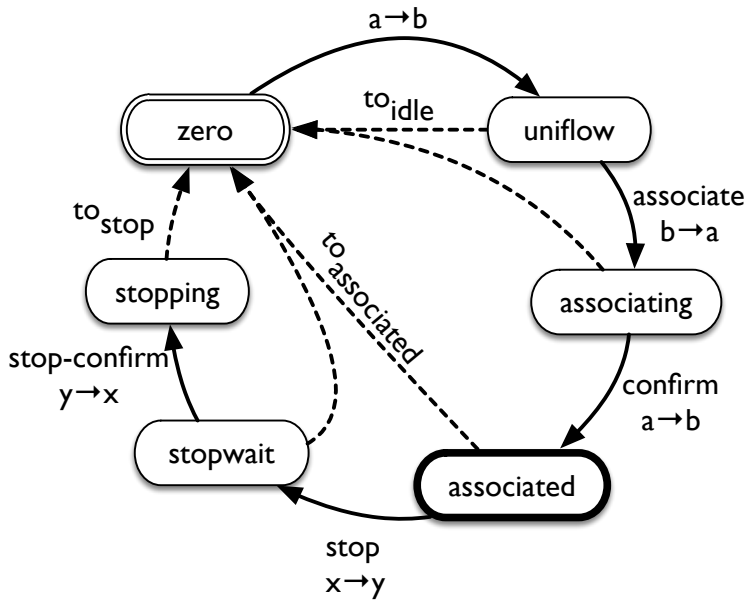

This state machine is primarily used to select among three timeouts (short *idle* and *stopping* timeouts, and a longer *associated* timeout) after which the on-path state for a flow will be dropped. Selection among these timeouts allows endpoints to reduce or eliminate unproductive keep-alive traffic needed to maintain on-path state and thereby end-to-end connectivity in the presence of NAT or other stateful on-path treatment.

The only possible active attack on the state signals lead to the premature selection of a shorter timeout. Careful design of the state signaling can mitigate this attack: notice in the state diagram above (the *stopwait* and *stopping* states) that both endpoints must send a stop signal before a device observing both directions of a flow adjusts its timeout down, requiring an attacker to have cooperating injection points on both sides of the on-path device.

Active attacks are in class (D, *).

## Treatment Signals

Packet Treatment Signals are signals from the sending endpoint to the path, informing on-path devices that a packet should be treated according to certain conditions.

PLUS for example has a Loss/Latency signal that says whether the packet is more sensitive to loss or more sensitive to latency. This is specified as a 1-bit sender-to-path signal that can be asserted on a per-packet basis. Its function is very similar to the one described in Latency Loss Tradeoff PHB Group. The main difference is that, in PLUS, the flag is included in the integrity protected envelope of the overlying transport, which makes it not as easily malleable as its DiffServ code-point counterpart.

We consider this signal as an example of other types of simple, tradeoff-based treatment signals.

From the perspective of a passive on-path observer, the Loss/Latency signal provides information about the class of traffic carried by a certain packet; namely, whether the packet can - or cannot - be associated with a source of latency sensitive traffic. This coarse categorization partitions the traffic space into two broad classes whose boundary can be, at times, a bit fuzzy. Examples of low-latency services include stock market flows to high frequency trading systems, real-time media applications (VoIP, interactive conferencing, etc.), game streaming services (e.g., Mixer, Twitch) and WebRTC-based low-latency live streaming. The latter is one example of an application that might interchangeably fall into one category or the other. (In fact, live streaming is often carried over an adaptive bit-rate bearer such as MPEG-DASH or HLS, which favours buffering over packet loss at queueing nodes.)

As described in Shbair et al. in A Multi-Level Framework to Identify HTTPS Services, state of art encrypted traffic analysis based machine learning can successfully identify the type of transported application (e.g., HTTPS, SMTP, P2P, VoIP, SSH, Skype) with good accuracy and without any need to access the clear-text. In this context, and despite its limitations (i.e., fuzzy, coarse grained), the Loss/Latency (LoLa) signal might be used to improve the precision of the classifier. This signal is non-malleable (can not be changed on-path). In contrast, the DiffServ field used to carry a DSCP, can be updated at a DiffServ edge router to map to available QoS treatments or to aggregate classes (including in some cases bleaching the field). The sender thus has no incentive to lie about the MCP LoLa marking, making it slightly more reliable signal compared using DiffServ.

Such attacks are in class (!D, !P).

### *Fingerprinting Attacks*

If a protocol provides (optional) capabilities to communicate with devices on a path either by requesting information from the path or providing information to the path, then an observer can use the patterns of this communication such as frequency of such communication and the kind of requested or provided information as a fingerprint. This could possibly allow the observer to reason about the application using the protocol. For example, applications that use such communication for high-precision RTT estimations can be identified from applications that do not require such estimations by observing whether requests for high-precision RTTs are observed.

The very presence of scratch spaces can also be used as a fingerprinting vector: an application or transport that uses scratch space reveals a little more of itself to the network. In other words, the pattern and frequency of requested PCFs could help identify the application.

# Risk Assessment and Conclusion

Looking at the previous sections in another way, we find the following attacks/features in classes (!D, !P), the most desirable class, or (!D, P), the next most desirable:

- Data exfiltration through scratch space (an active attack)

- Connection identifier

- Packet treatment signals

- Packet number (as a diminished form of connection identifier)

*In general, middlebox cooperation does not enable any new classes of attacks beyond what is already possible with standard TCP or UDP and encrypted payload.*

The active attack of data exfiltration through scratch space can be performed by any two colluding middleboxes simply by one middlebox sending datagrams to the other. Network monitoring may detect these packets, however, it is currently unusual for two middleboxes to talk to each other directly.

The passive attacks on a connection identifier can be performed today by middleboxes observing five-tuples. Giving an explicit connection identifier will make the job of the passive attacker easier. Packet treatment signals such as loss/latency can already be inferred from traffic characteristics, transport protocol port numbers, or endpoint address.

With TCP, packet numbers are implicitly present and also carried in acknowledgment numbers. In contrast, the UDP protocol header does not specify any sequence numbers - although protocols layered on top of UDP may introduce these (e.f., RTP, SCTP, DCCP). Other network-layer encapsulations may permit the use of sequence numbers (e.g. in a GRE option field), but do not normally have a sequence number field that is observable by devices on the path.

An MCP, no matter how it may be constructed, will most likely have information with fixed formats and meaning, as well as free-form information. Information with fixed formats can be abused for data exfiltration and linkability, but under the assumption that such data is integrity-protected, and under the requirement that normal operation should not be disrupted too much, it should be restored to a close-to-original state before being delivered to the receiver. That limits the abuse potential of that data.

The fixed-form data present in an MCP protocol header are of course dual-use: for any legitimate use of MCP header data, there is an equal and opposite abuse. For example, a loss/latency indicator could be used to distinguish real-time traffic from file downloads; start and stop bits could be used to delimit flows, and so on. The opportunities for this kind of abuse are however not different from the possibilities present in TCP.

The potential for new abuse comes from introducing free-form data, which seems to present a practically unlimited amount of scratch space with which an attacker could exfiltrate data. On the surface, this seems to provide a potential for attacks against privacy. But network operators can encapsulate packets or add tunnel headers, or set tags in existing encapsulations. They can also export flow information that allows other devices to measure the traffic or react in a particular way. These techniques are readily applied within a single operator domain, but become more difficult to coordinate across network boundaries.

To use scratch space for intra-network exfiltration, the adversary has to control a portion of the path through which the packets travel. Such an adversary could also use other means to accomplish the same effect, without even changing the original packets. One easy way would be for an adversary to send control datagrams to an on-path device. A network operator therefore has many more tools to help support the service that is required from their customers. IP destination address is often a pretty helpful indicator to determining pattern of use.

Middlebox cooperation may make the attacker's job easier in some cases, for example by giving it information explicitly that it would otherwise have to work to get, by making classifiers more accurate, and so on. However, in general, middlebox cooperation does not enable any new classes of attacks beyond what is already possible with standard TCP or UDP and encrypted payload.

This work is partially supported by the European Commission under Horizon 2020 grant agreement no. 688421 Measurement and Architecture for a Middleboxed Internet (MAMI), and by the Swiss State Secretariat for Education, Research, and Innovation under contract no. 15.0268. This support does not imply endorsement.